\documentclass{article}
\usepackage{graphics,graphicx,array,hhline,color,float,mdwlist,amsmath,amssymb,cite,multirow,xcolor,hyperref,tabularx,placeins,xspace}
\usepackage[left=2cm,right=2cm,top=2cm,bottom=2cm,head=1cm,foot=1cm]{geometry}
\graphicspath{{./figs/}}
\usepackage{t1enc}
\usepackage[utf8]{inputenc}

\newcommand{\m}[1]{\mathrm{#1}}

\newcommand{\mT}{m_{T}}
\newcommand{\pT}{p_{T}}

\begin{document}
\title{L\'evy femtoscopy with PHENIX at RHIC}
\author{M\'at\'e Csan\'ad for the PHENIX Collaboration\\
E\"otv\"os Lor\'and University, H-1117 Budapest, P\'azm\'any P. s. 1/A, Hungary}
\maketitle

%\keyword{RHIC, PHENIX, femtoscopy, Bose-Einstein correlations, L\'evy distribution, anomalous diffusion, critical point, in-medium mass modification}

\begin{abstract}In this paper we present the measurement of charged pion two-particle femtoscopic correlation functions
in $\sqrt{s_{NN}}=200$ GeV Au+Au collisions, in 31 average transverse mass bins, separately for positive and
negative pion pairs. L\'evy-shaped source distributions yield a statistically acceptable description of the measured
correlation functions, with three physical parameters: correlation strength parameter $\lambda$, L\'evy index
$\alpha$ and L\'evy scale parameter $R$. The transverse mass dependence of these L\'evy parameters is
then investigated and their physical interpretation is also discussed, and the appearance of a new scaling variable is observed.
\end{abstract}

\section{Introduction}
In collisions at the Relativistic Heavy Ion Collider, a strongly coupled Quark Gluon Plasma is
formed~\cite{Adcox:2004mh,Adams:2005dq,Arsene:2004fa,Back:2004je}, creating hadrons
at the freeze-out. The measurement of femtoscopic correlation functions is used to infer the 
space-time extent of hadron creation. The field of femtoscopy was founded by the astronomical
measurements of R. Hanbury Brown and R. Q. Twiss~\cite{HanburyBrown:1956bqd}
and the high energy physics measurements of G. Goldhaber and collaborators~\cite{Goldhaber:1959mj,Goldhaber:1960sf}.
If interactions between the created hadrons, higher order correlations, decays and all other dynamical two-particle correlations
may be neglected, then the two-particle Bose-Einstein correlation function is simply related to the source function $S(x,k)$
(which describes the probability density of particle creation at the space-time point $x$ and with
four-momentum $x$). This can be understood if one defines
$N_1(p)$ as the invariant momentum distribution and $N_2(p_1,p_2)$ as the momentum pair distribution.
Then the definition of the correlation function is~\cite{Yano:1978gk}:
\begin{align}
C_2(p_1,p_2)&=\frac{N_2(p_1,p_2)}{N_1(p_1)N_1(p_2)},\textnormal{ where }\\
N_2(p_1,p_2)&=\int S(x_1,p_1)S(x_2,p_2)|\Psi_2(x_1,x_2)|^2d^4x_2d^4x_1.
\end{align}
Here $\Psi_2(x_1,x_2)$ is two-particle wave function, for which
\begin{align}
\left|\Psi_2(x_1,x_2)\right|^2=\left|\Psi^{(0)}_2(x_1,x_2)\right|^2=1+\cos(p_1-p_2)(x_1-x_2)
\end{align}
follows in an interaction-free case. This leads to 
\begin{align}
C^{(0)}_2(Q,K)&\simeq 1+\left|\frac{\widetilde{S}(Q,K)}{\widetilde{S}(0,K)}\right|^2,\;\;\textnormal{where}\label{e:C0}\\
 \widetilde{S}(q,k)&=\int S(x,k)e^{iqx}d^4x\;\;\textnormal{is the Fourier-transformed of } S,
\end{align}
and $Q = p_1 - p_2$ is the momentum difference, $K = (p_1+p_2)/2$ is the average
momentum, and we assumed, that $Q \ll K$ holds for the investigated kinematic range.
Usually, correlation functions are measured versus $Q$, for a well-defined $K$-range,
and then properties of the correlation functions are analyzed as a function of the average $K$
of each range. If the source is a static Gaussian with a radius $R$, then the correlation function will also
be a Gaussian with an inverse radius, hence it can be described by one plus a 
Gaussian: $1+\exp -(qR)^2$. However, if the source is expanding, then the observed
Gaussian radius $R$ does not represent the geometrical size, but rather a length of
homogeneity, depending on the average momentum $K$. The approximate dependence
of  $R^{-2}\propto a+b \mT$ is observed for various collision systems,
collision energies and particle types~\cite{Adler:2004rq,Afanasiev:2009ii}, which 
can be interpreted as a consequence of hydrodynamical expansion~\cite{Makhlin:1987gm,Csorgo:1995bi}.
See Ref.~\cite{Adare:2017vig} (and references therein) for details.

Usually,  the shape of the particle emitting source is assumed to be Gaussian,
however, this does not seem to be the case experimentally~\cite{Afanasiev:2007kk,Adler:2006as}.
In an expanding hadron resonance gas, increasing mean free paths lead to a L\'evy-flight,
anomalous diffusion, and hence to spatial L\'evy distributions~\cite{Metzler:1999zz,Csorgo:2003uv,Csanad:2007fr}.
The one-sided, symmetric L\'evy distribution is defined as:
\begin{align}
\mathcal{L}(\alpha,R,r)=(2\pi)^{-3} \int d^3q e^{iqr} e^{-\frac{1}{2}|qR|^{\alpha}},
\end{align}
where $\alpha$ is the L\'evy index and $R$ is the L\'evy scale. Then $\alpha=2$ gives back the Gaussian
case and $\alpha=1$ yields a Cauchy distribution. This source function leads to a correlation function of
\begin{align}
C_{2}(Q,K)=1+\lambda\cdot e^{-(Q\cdot R(K))^{\alpha}}.
\end{align}
It is interesting to observe that the spatial L\'evy distribution results in 
power-law tails in the spatial correlation function, with an exponent of $-1-\alpha$.
Such power-law spatial correlations are also expected in case of critical behavior,
with an exponent of $-(d-2+\eta)$, with $\eta$ being the critical exponent.
It is easy to see that in this case, $\eta=\alpha$, i.e. the L\'evy exponent
is identical to the critical exponent $\eta$~\cite{Csorgo:2009gb}.
The second order QCD phase transition is expected to be in the same
universality class as the phase transition of the 3D Ising model (see
Refs~\cite{El-Showk:2014dwa,Rieger:1995aa,Halasz:1998qr,Stephanov:1998dy}
for details and values for the critical exponents), and hence around the
critical point, $\alpha\leq 0.5$ values may be expected~\cite{Csorgo:2009gb}.
Since the exploration of the QCD phase diagram, in particular the search for
the QCD critical endpoint is one of the major goals of experimental heavy ion physics
nowadays, the above discussed relations yield additional motivation for the measurement
and analysis of of Bose-Einstein correlation functions.

Furthermore, it is important to note, that not all pions are primordial, i.e. not all of them are created
in directly from the collision. A significant fraction of pions are secondary, coming from decays.
Hence the source will have two components: a core of primordial pions, stemming from the
hydrodynamically expanding sQGP (and the decays of very short lived resonances, with half-lives less
than a few fm$/c$), and a halo, consisting of the decay products of long lived resonances
(such as $\eta$, $\eta'$, $K^0_S$, $\omega$)
\begin{align}
S=S_{\rm core}+S_{\rm halo}
\end{align}
These two components have characteristically different sizes ($\lesssim 10$ fm for the core,
$>50$ fm for the halo, based on the half-lives of the above mentioned resonances).
In particular, the halo component is so narrow in momentum-space,
that it cannot be resolved experimentally. This leads to
\begin{align}\label{ecorehalo}
C^{(0)}_2(Q,K)&=1+\frac{|\widetilde S(Q,K)|^2}{|\widetilde S(0,K)|^2}\simeq
1+\left(\frac{N_\m{core}(K)}{N_\m{core}(K)+N_\m{halo}(K)}\right)^2
\frac{|\widetilde S_{\rm core}(Q,K)|^2}{|\widetilde S_{\rm core}(0,K)|^2}\\
&=1+\lambda(K)\frac{|\widetilde S_{\rm core}(Q,K)|^2}{|\widetilde S_{\rm core}(0,K)|^2}\nonumber
\end{align}
where $N_\m{core}(K)=\int S_{\rm core}(Q,K) dQ$ and $N_\m{halo}(K)=\int S_{\rm halo}(Q,K) dQ$
were intruced. Furthermore 
\begin{align}
\lambda(K) = \left(\frac{N_\m{core}(K)}{(N_\m{core}(K)+N_\m{halo}(K)}\right)^2,
\end{align}
was defined, equivalent to the ``intercept'' of the correlation function:
\begin{align}
\lim\limits_{Q\rightarrow 0} C^{(0)}_2(Q,K) = 1 + \lambda(K).
\end{align}
Hence, in the core-halo picture, $\lambda(K)$ is related to the fraction of primordial (core)
pions among all (core plus halo) pions at a given momentum. One of the motivations for measuring
$\lambda$ is that it is related~\cite{Vance:1998wd} to the $\eta'$ meson yield,
expected~\cite{Kapusta:1995ww} to increase in case of chiral $U_A(1)$ symmetry
restoration in heavy-ion collisions (due to the expected in-medium mass decrease of the $\eta'$).

We also have to take into account that the interaction-free case is not valid for the usual measurement of
charged particle pairs, the electromagnetic and strong interactions distort the above simple picture.
For identical charged pions, the Coulomb interaction is the most important, and this decreases the number of particle
pairs at low momentum differences. This can be taken into account by utilizing the
$\Psi^{(C)}_2(x_1,x_2)$ pair wave function solving the Schr\"odinger-equation for charged particles,
given for example in Ref.~\cite{Adare:2017vig}. With this, a so-called ``Coulomb-correction'' can be calculated as
\begin{align}\label{e:KCoulomb}
K_2(Q,K) = 
\frac{
\int d^4x_1 d^4x_2 S(x_1,K-Q/2) S(x_2,K+Q/2) \left|\Psi^{(C)}(x_1,x_2)\right|^2}{
\int d^4x_1 d^4x_2 S(x_1,K-Q/2) S(x_2,K+Q/2) \left|\Psi^{(0)}(x_1,x_2)\right|^2},
\end{align}
and using this, the measured correlation function can be described by
\begin{align}\label{e:KCoulomb}
C_2^{\rm measured}(Q,K) = K_2(Q,K)C^{(0)}_2(Q,K).
\end{align}
For details, see again Ref.~\cite{Adare:2017vig} and references therein.

In the following, we utilize a generalization of the usual Gaussian shape of
the Bose-Einstein correlations, namely we analyze our data using L\'evy stable
source distributions. We have carefully tested that this source model is in
agreement with our data in all the transverse momentum regions reported here:
all the L\'evy fits were statistically acceptable, as discussed also later. 
We note that using the method of L\'evy expansion of the
correlation functions~\cite{Novak:2016cyc}, we have found that within errors all the
terms that measure deviations from the L\'evy shape are consistent with zero.
Hence we restrict the presentation of our results to the analysis of the
correlation functions in terms of L\'evy stable source distributions.

\begin{figure}
\begin{center}
\includegraphics[width=0.6\textwidth]{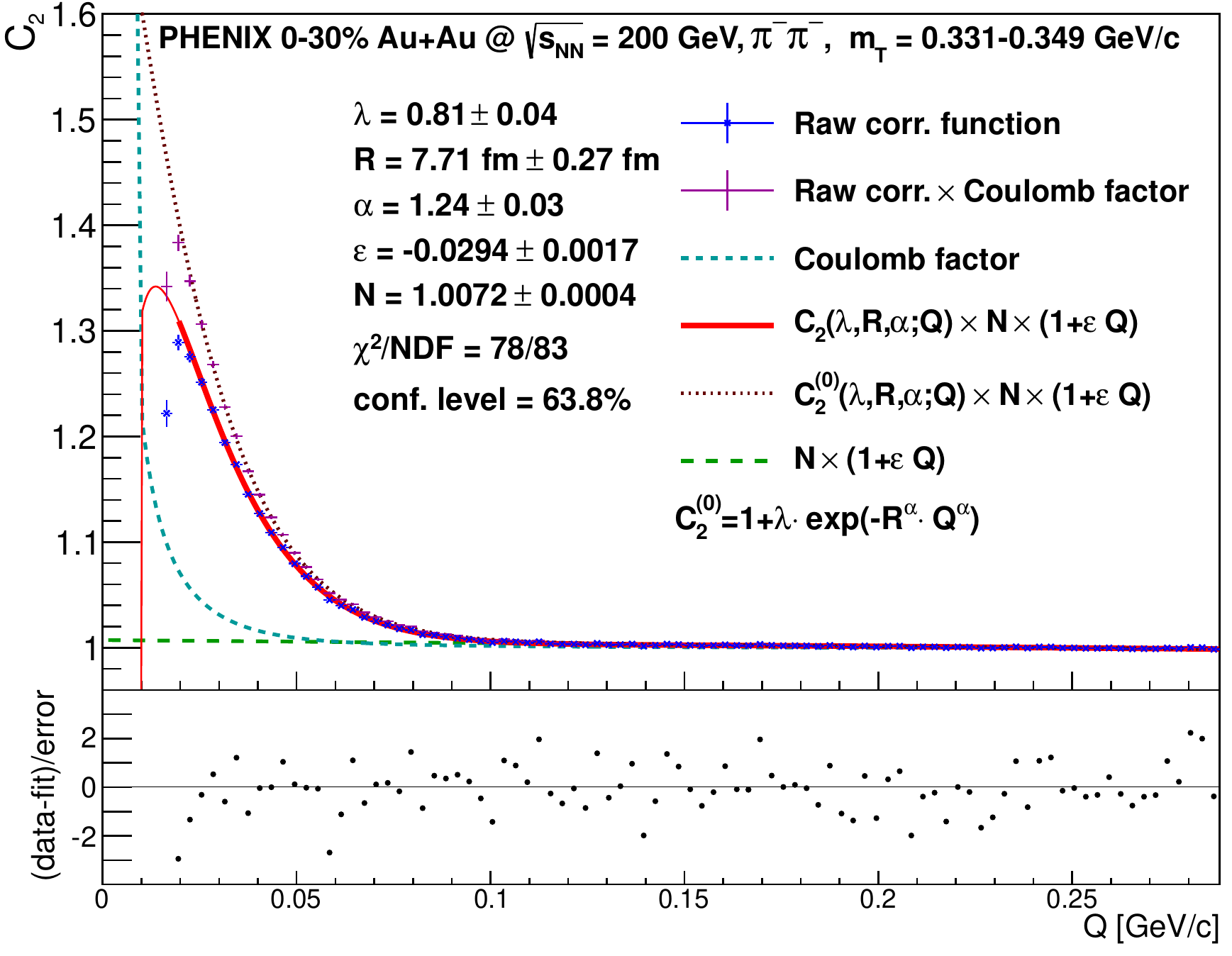}
\end{center}
\caption{Example fit of to $\pi^+\pi^+$ pairs with average $\pT$ between 0.2 and 0.22 GeV/c,
measured in the longitudinal co-moving frame. The fit shows the measured correlation function and the complete
fit function, while a ``Coulomb-corrected" fit function $C^{(0)}(k)$ is also shown, with the data multiplied
by $C^{(0)}/C^{\rm Coul}$.} \label{f:fit}
\end{figure}

\section{Results}

\begin{figure}
\begin{center}
\includegraphics[width=0.4\textwidth]{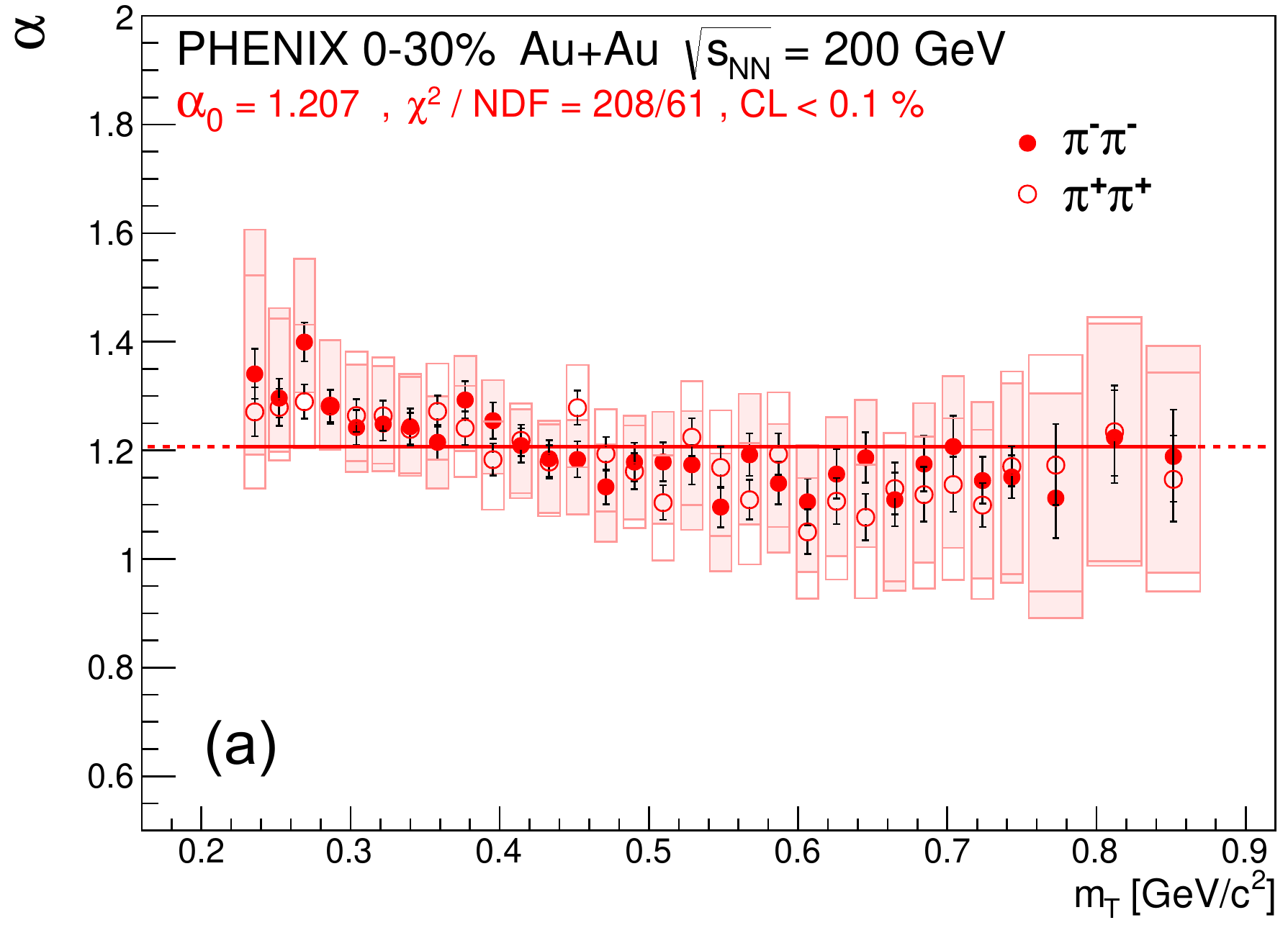}
\includegraphics[width=0.4\textwidth]{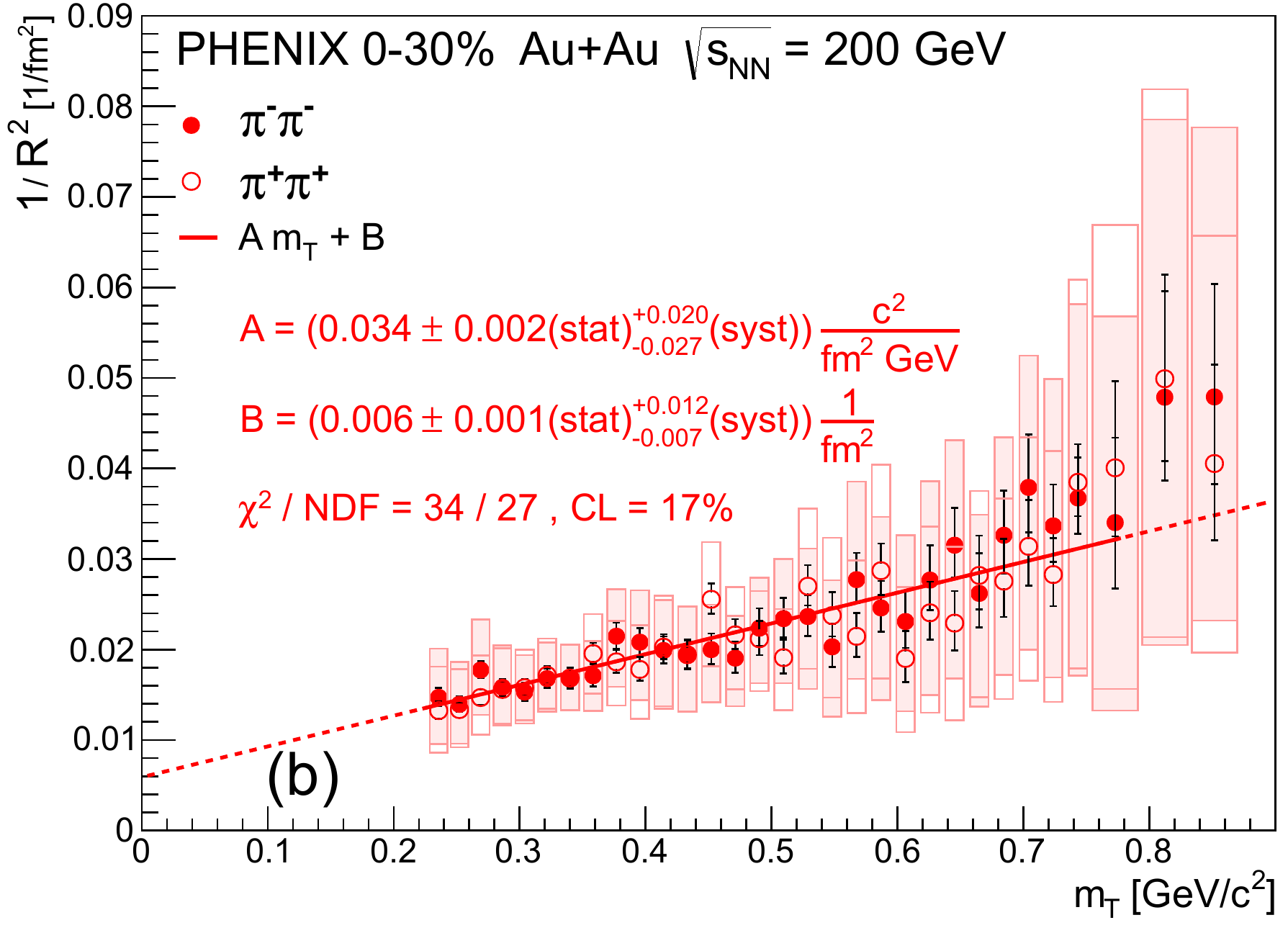}\\
\includegraphics[width=0.4\textwidth]{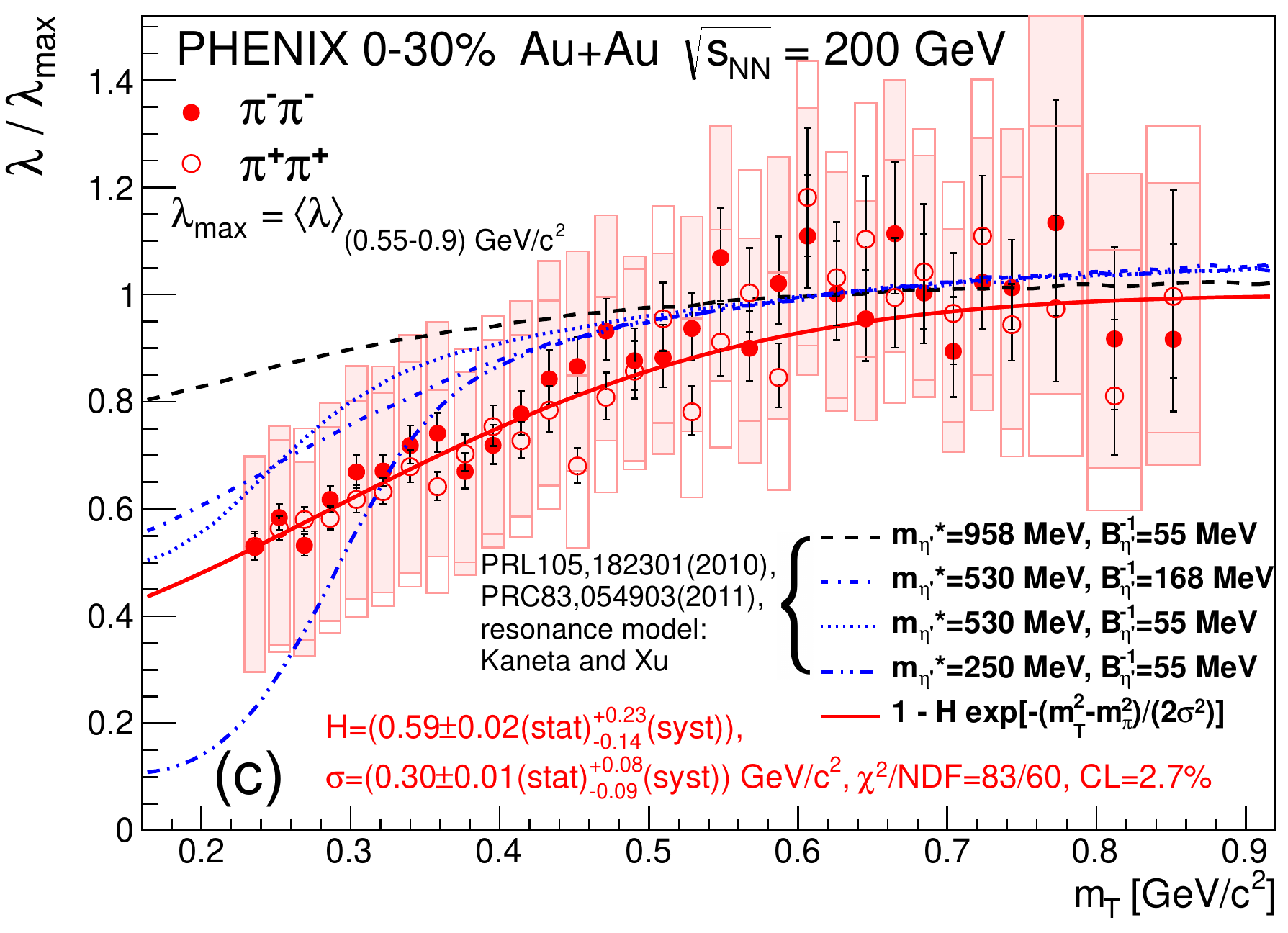}
\includegraphics[width=0.4\textwidth]{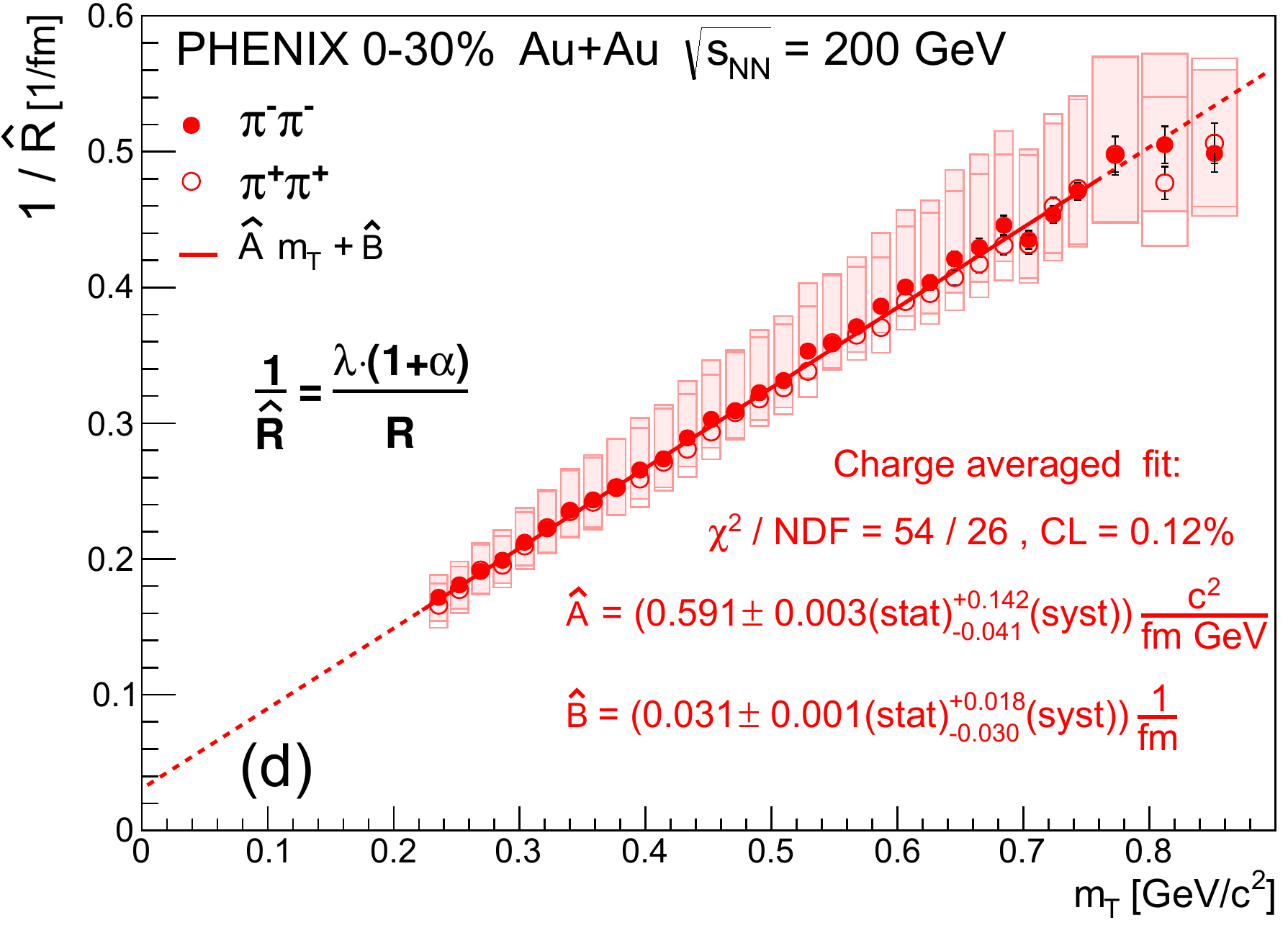}
\end{center}
\caption{Fit parameters $\alpha$ (top left), $R$ (top right, shown as $1/R^2$), $\lambda$ (bottom left, shown as
$\lambda/\lambda_{\rm max}$) and $\widehat{R}$ versus average $m_T$ of the pair. Statistical and symmetric
systematic uncertainties shown as bars and boxes, respectively.}
\label{f:results}
\end{figure}

We analyzed 2.2 billion 0--30\% centrality $\sqrt{s_{NN}} = $ 200 GeV Au+Au collisions recorded by
the PHENIX experiment in the 2010 running period.\footnote{In the conference presentation Minimum Bias data
were shown, but here we show and discuss the final  0--30\% centrality data of Ref.~\cite{Adare:2017vig}.
The Minimum Bias data are available e.g. in Ref.~\cite{Kincses:2016jsr}.} We measured two-particle correlation functions
of $\pi^-\pi^-$ and $\pi^+\pi^+$ pairs, in 31 $\mT$ bins ranging from 228 to 871 MeV, as detailed
in Ref.~\cite{Adare:2017vig}. The calculated correlation functions based on L\'evy-shaped sources
gave statistically acceptable descriptions of all measured correlation functions (all transverse momenta and both 
charges). This allow us to study and interpret the $\mT$ dependence of the fit parameters  $R$, $\alpha$ and
$\lambda$, as they do represent the measured correlation functions. 

Let us turn to the resulting fit parameters and their $\mT$ dependence. Parameters $\lambda$, $R$ and $\alpha$ are
shown in Fig.~\ref{f:results}, as a function of pair $\mT$ (corresponding to the given $\pT$ bin). The detailed
description of the systematic uncertainties is given in Ref.~\cite{Adare:2017vig}, here we focus on the
characteristics of the $\mT$ dependencies. In the top left panel of Fig.~\ref{f:results}, we observe
that $\alpha$ is constant within systematic uncertainties, with an average value of 1.207. This average $\alpha$
value is far from the Gaussian assumption of $\alpha=2$ as well as from the conjectured $\alpha=0.5$ value at the critical point.
We show $1/R^2$ as a function of $\mT$ on the top right panel of Fig.~\ref{f:results}. We observe, that the hydro prediction
of  $1/R^2\simeq a+b\mT$ still holds -- even though in hydrodynamics,
usually no power-law tails appear, due to the Boltzmann factor creating an exponential cut-off.
This is intriguing point may be investigated in phenomenological models in the future.
The correlation function intercept parameter $\lambda$ is shown in the bottom left panel of Fig.~\ref{f:results},
after a normalization by
\begin{align}
\lambda_{\rm max}=\langle \lambda \rangle_{\mT=0.5-0.7 {\rm GeV}/c^2},
\end{align}
as detailed in Ref.~\cite{Adare:2017vig}.
The $\mT$ dependence of $\lambda/\lambda_{\rm max}$ indicates a decrease at small $\mT$. This
may be explained by the increase of the resonance pion fraction at low $\mT$. Such an 
increase is predicted to occur in case of an in-medium $\eta'$ mass, as discussed above. It is
interesting to observe that our data are not incompatible with predictions based on a reduced $\eta'$ mass.
Finally, in the bottom right panel of Fig.~\ref{f:results} we show the observation of a new scaling parameter
\begin{align}
\widehat{R}= \frac{R}{\lambda(1+\alpha)}.
\end{align}
The inverse of this variable exhibits a clear linear scaling with $\mT$, and it also has much decreased statistical uncertainties. Let us conclude
by inviting the theory/phenomenology community to calculate the $\mT$ dependence of the above
L\'evy parameters, and compare their result to the measurements.

\section*{Acknowledgments}
M. Cs. was supported by the New National Excellence program of the Hungarian Ministry of Human Capacities, the NKFIH grant FK-123842 and the J\'anos Bolyai Research Scholarship.

\bibliographystyle{../../../prlsty}
\bibliography{../../../Master}

\end{document}